\documentclass{aa501}
\usepackage{graphicx}
\def\ltsim{\raise 2pt \hbox {$<$} \kern-1.1em \lower 4pt \hbox {$\sim$}}
\def\ltapprox{\raise 2pt \hbox {$<$} \kern-1.1em \lower 5pt \hbox {$\approx$}}
\def\gtsim{\raise 2pt \hbox {$>$} \kern-1.1em \lower 4pt \hbox {$\sim$}}
\def\gtapprox{\raise 2pt \hbox {$>$} \kern-1.1em \lower 5pt \hbox {$\approx$}}

\def\arcsec{$^{\prime\prime}$}
\def\arcmin{$^{\prime}$}
\def\degrees{$^{\circ}$}
\def\etal{{et al.~}}
\begin{document}
%
\title
{The giant radio halo in Abell 2163 }
\author{L. Feretti\inst{1} 
\and R. Fusco-Femiano\inst{2} 
\and G. Giovannini\inst{1,3} 
\and F. Govoni\inst{1,4}}
\offprints{L. Feretti}
\institute{
Istituto di Radioastronomia -- CNR, via P. Gobetti 101, I--40129
Bologna, Italy
\and Istituto di Astrofisica Spaziale CNR, via del Fosso del Cavaliere,
I--00133 Roma, Italy
\and Dipartimento di Fisica, Univ. Bologna,  
Via B. Pichat 6/2, I--40127 Bologna, Italy
\and Dipartimento di Astronomia, Univ. Bologna, via Ranzani 1, 
I-40127 Bologna, Italy}

   \date{Astronomy \& Astrophysics, in press}

\abstract{
New radio  data is presented for the rich cluster Abell 2163.
The cluster radio emission is characterized by the presence of a
 radio halo, which is one
of the most powerful and extended halos known so far.
In the NE peripheral cluster region, we also detect diffuse elongated 
emission, which we classify as a cluster relic.
The cluster A2163 is very hot and luminous in X-ray. Its central region
is probably in a highly non relaxed state, suggesting that this
cluster is likely to be a recent merger. The existence of a radio halo
in this cluster confirms that halos are associated with hot massive
clusters, and confirms the connection between radio halos and 
cluster merger processes.
The comparison between the radio emission of the halo and 
the cluster X-ray emission shows a close structural similarity. 
A power law correlation is found
between the radio and X-ray brightness, with index = 0.64.
We also report the upper limit to the hard X-ray emission,
obtained from a BeppoSAX observation.
We discuss the implications of our results. 
}

\maketitle

\section{Introduction}

Recent observations of clusters of galaxies have revealed a new and complex
scenario in the structure of the intergalactic medium. 
The clusters are not simple
relaxed structures, but are still forming at the present epoch. Substructures,
commonly observed  in the X-ray distribution of  a high number of rich
clusters (Henry \& Briel 1993, Burns \etal 
1994),  are evidence of the hierarchic growth of clusters from the merger
of poorer subclusters.

An important problem in cluster phenomenology involves cluster-wide radio
halos, whose prototype is Coma C (Giovannini \etal 1993, and references
therein). These are extended diffuse radio sources  located at the
cluster centers, with typical sizes of
\gtsim 1 $h_{50}$ Mpc, regular shape, steep radio spectra
and no significant polarization. 
Diffuse radio sources, named relics,  have also been detected  
in peripheral regions of the clusters. They are irregular in shape and
generally highly polarized. The
origin and properties of halos and relics are still being 
debated. According to
recent suggestions, the cluster merger process
may play a crucial role in
the formation and energetics of these sources
(see Feretti \& Giovannini 1996 and references therein, Feretti 1999). 

Abell 2163
is a distant (z=0.203, Struble \& Rood, 1999),
rich cluster, and is one of the hottest 
(mean kT = 12-15 keV, Elbaz \etal 1995, Markevitch \etal 1996),
and most X-ray luminous (L$_{\rm X[2-10keV]}$ = 6 $\times$ 10$^{45}$
erg s$^{-1}$, Arnaud \etal 1992) among known  clusters. A BeppoSAX observation
measures a temperature of 10-11 keV out to $4^{\prime}$, with a marginally 
significant ($<2\sigma$) rise in temperature at larger radii 
(Irwin \& Bregman 2000).
According to the morphological study of Elbaz \etal (1995),
based on ROSAT data, and the spectroscopic analysis 
of Markevitch \etal (1996), based on ASCA data,
the cluster is likely to be a recent merger. A highly nonrelaxed state
in the cluster inner region is  confirmed
by the recent Chandra results (Markevitch \etal 2000). 
A strong  Sunyaev-Zel'dovich (SZ) effect has been reported
in this cluster (Holzapfel \etal 1997, D\'esert \etal 1998).  

We present here the radio data of A2163 obtained with the Very Large Array
(VLA). In particular,
we analyze in detail the diffuse radio emission whose presence was reported
by Herbig \& Birkinshaw (1994).
We also give data for the extended radio galaxies.
In addition, we report the results obtained by the Phoswich Detector
System (PDS) on board BeppoSAX to search for hard X-ray radiation.

We adopt H$_0$=50 km s$^{-1}$ Mpc$^{-1}$ and q$_0$ = 0.5. With these
values, 1 arcsec corresponds to 4.27 kpc at the distance of A2163. 

\section{Radio Observations}

Radio observations  were obtained with the VLA
in different configurations (see Table 1) at 20 cm, and with the D array
at 6 cm. The
data were calibrated and
reduced with the  Astronomical Image Processing
System (AIPS), 
following the standard procedure (Fourier inversion,
CLEAN and RESTORE, selfcalibration). 

\begin{figure} 
\vspace{6 cm}
\caption{ Radio map at 20 cm with resolution of 45\arcsec$\times$60\arcsec
(RA$\times$DEC).
The $\sigma$ noise level in this map is 0.03 mJy/beam.
Contours are at  --0.1, 0.1, 0.2, 0.3, 0.5, 0.7, 1, 2, 3, 5, 10, 25 mJy/beam.
Crosses indicate the discrete radio sources detected from the higher
resolution images. }
\end{figure}

\begin{table}
\caption{Observing log}
\begin{flushleft}
\begin{tabular}{lllllll}
\hline
\noalign{\smallskip}
Freq. &  Bandwitdh  & Array  &  Date  & Duration \\
  Mhz & MHz  & & & min \\
\noalign{\smallskip}
\hline
\noalign{\smallskip}
 1365/1465 & 50 & D & 02 FEB 1998 & 140 \\
 1365/1465 & 25 & C & 02 JAN 1999 & 200 \\
 1365/1465 & 25 & A & 21 MAY 1998 & 50 \\
 4885/4835 & 50  & D & 26 DEC 1997 & 120 \\
\noalign{\smallskip}
\hline
\noalign{\smallskip}
\end{tabular}
\end{flushleft}
\end{table}

At 20 cm, images with angular resolutions ranging from $\sim$1.5\arcsec~ to
60\arcsec~ were produced.
At the lowest resolution, maps
for the two observing frequencies were produced separately, to get some
information about the spectral index.
We also obtained maps of the polarized intensity in the standard way.
An image at 6 cm was produced with resolution 15\arcsec$\times$20\arcsec
(RA$\times$DEC), and a rms noise level of 0.35 mJy/beam.

\section{Results}

\subsection{Radio halo and other diffuse sources} 

The radio image of this cluster  is shown in 
Fig. 1, where the crosses mark discrete unrelated sources
detected in the map at 4\arcsec~ resolution, with a 
brightness higher than 0.2 mJy/beam.
The diffuse emission permeates the cluster center for  a
total extent of $\sim$ 11.5\arcmin~, corresponding 
to $\sim$2.9  Mpc. With this size, the halo in A2163
is the largest known radio halo. It displays a very  regular shape,
slightly elongated in the E-W direction. The low brightness
emission does not fade into the noise, but shows rather sharp boundaries.
This is an indication that the whole structure is properly imaged
here.
We made polarization maps of the cluster at 20 cm. 
Some polarized flux is detected at a level of 2-4\%. Because of the
noise, this should be considered as an upper limit.

\begin{figure}
\vspace{6cm}    
\caption{ Radio map at 20 cm with resolution of 15\arcsec, overlayed
onto the grey-scale image from the digitized Palomar Sky Survey.
The $\sigma$ noise level in this map is 0.03 mJy/beam.
Contours are at  --0.09, 0.09, 0.15, 0.3, 0.6, 1.2, 2.5 mJy/beam.
Labels indicate the tailed radio galaxies and  the diffuse
features (see text).}
\end{figure}

At higher angular resolution (Fig. 2), only the central region
of the radio halo is detected. All the discrete sources embedded within
the diffuse emission are easily visible.
The high resolution image reveals the existence of diffuse features 
at the periphery of the cluster (labeled D1, D2, D3 and D4 in Fig. 2). 
The diffuse emission D1 seems to be coincident with a faint
galaxy. The northern source D2 could be related to a nearby 
point-like
source. The two emissions D3 and D4 do not show any obvious
optical counterpart. If they 
were both associated with  the strong eastern radio
source at position RA = 16$^{\rm h}$ 16$^{\rm m}$ 22.2$^{\rm s}$, 
DEC = --06\degrees~ 06\arcmin~ 
34\arcsec, which coincides with a faint galaxy, they  would form 
a wide-angle-tailed (WAT) radio source of about 1.3 Mpc in 
size (assuming it is a cluster 
radio galaxy). This seems unlikely,  since the size would 
be quite large for 
a WAT (e.g., O'Donoghue \etal 1990) and since 
WAT sources are usually located at the cluster center.
 Although the above arguments may be weak,  we suggest the  
reasonable possibility that at least the source D3 is a 
diffuse peripheral relic, similar to the source 1253+275 in the Coma cluster
(Giovannini \etal 1991), while the source D4 could be connected either to the 
mentioned
point source or to the relic. More observations are necessary to investigate 
this point.

\begin{table}
\caption{Diffuse sources}
\begin{flushleft}
\begin{tabular}{lll}
\hline
\noalign{\smallskip}
 & Halo &  Relic (D3) \\
\noalign{\smallskip}
\hline
\noalign{\smallskip}
Flux Density - 20 cm (mJy)                    & 155$\pm$2  &  18.7$\pm$0.3 \\
Largest Size (kpc)                            & 2900$\pm$100 &  640$\pm$50 \\
Spectral index                                & 1.6$\pm$0.3 & 2.1$\pm$0.3  \\
Power - 20 cm ($\times$10$^{24}$ W Hz$^{-1}$) & 30.1$\pm$0.4 & 3.63$\pm$0.06  \\ 
Luminosity ($\times$10$^{35}$ W)              &  13.6$_{-7.8}^{+26.9}$ 
& 10.6$_{-7.3}^{+26.6}$    \\
Volume ($\times$10$^{7}$ kpc$^3$)             &  $\sim$950 & $\sim$3.2 \\
u$_{min}$ ($\times$10$^{-14}$ erg cm$^{-3}$)  &  8.2 & 36 \\
U$_{min}$ ($\times$10$^{61}$ erg)             &  2.3 & 0.34 \\  
H$_{eq}$ ($\mu$G)                             &  0.9 &  2.0 \\
\noalign{\smallskip}
\hline
\noalign{\smallskip}
\end{tabular}
\end{flushleft}
\end{table}

\begin{figure}
\includegraphics{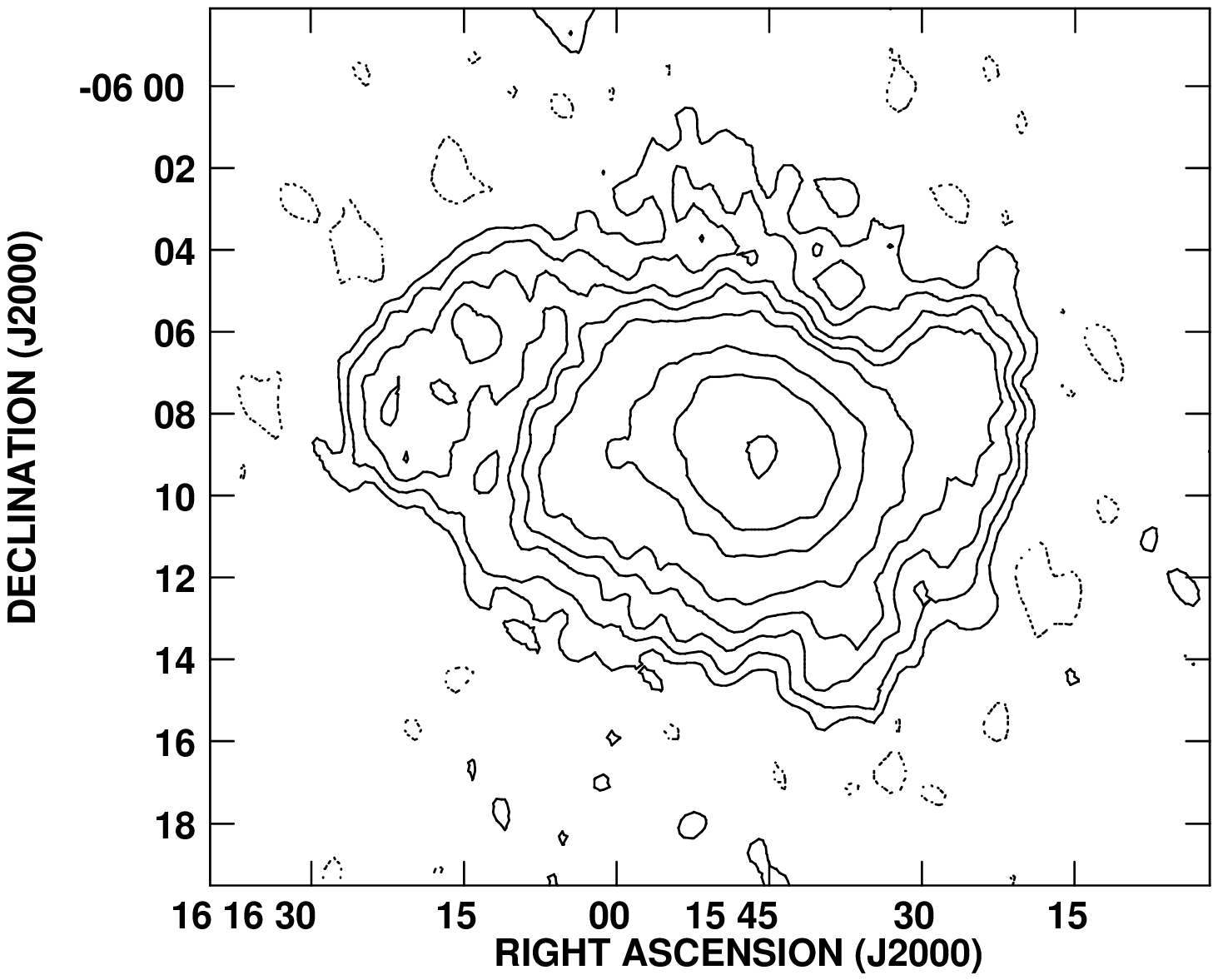}
\vspace{9 cm}
\caption{ Radio map at 20 cm of the diffuse emission, after 
subtraction of the discrete unrelated sources. The resolution is 
the same as in Fig. 1, whereas the noise level is 0.05 mJy/beam.
Contour levels are -0.15, 0.15, 0.3, 0.5, 0.7, 1, 2, 3, 5 mJy/beam.
}
\end{figure}

We obtained a 20 cm image of the radio halo 
after subtraction of the discrete unrelated sources.
To this aim, we produced an image of discrete sources,
by selecting from the original data set only
the data from the long baselines, which do not contain the 
extended halo emission. This image was then subtracted from the 
image of Fig. 1. The halo map obtained in this way is presented  
in Fig. 3. The diffuse emission 
surrounding the halo is due to the relic and to the other
diffuse features, which are extended and therefore difficult
to be subtracted entirely. 
The parameters for the central radio halo, after subtraction of the 
unrelated sources,  and for the relic 
are presented in Table 2.

\begin{figure}
\includegraphics{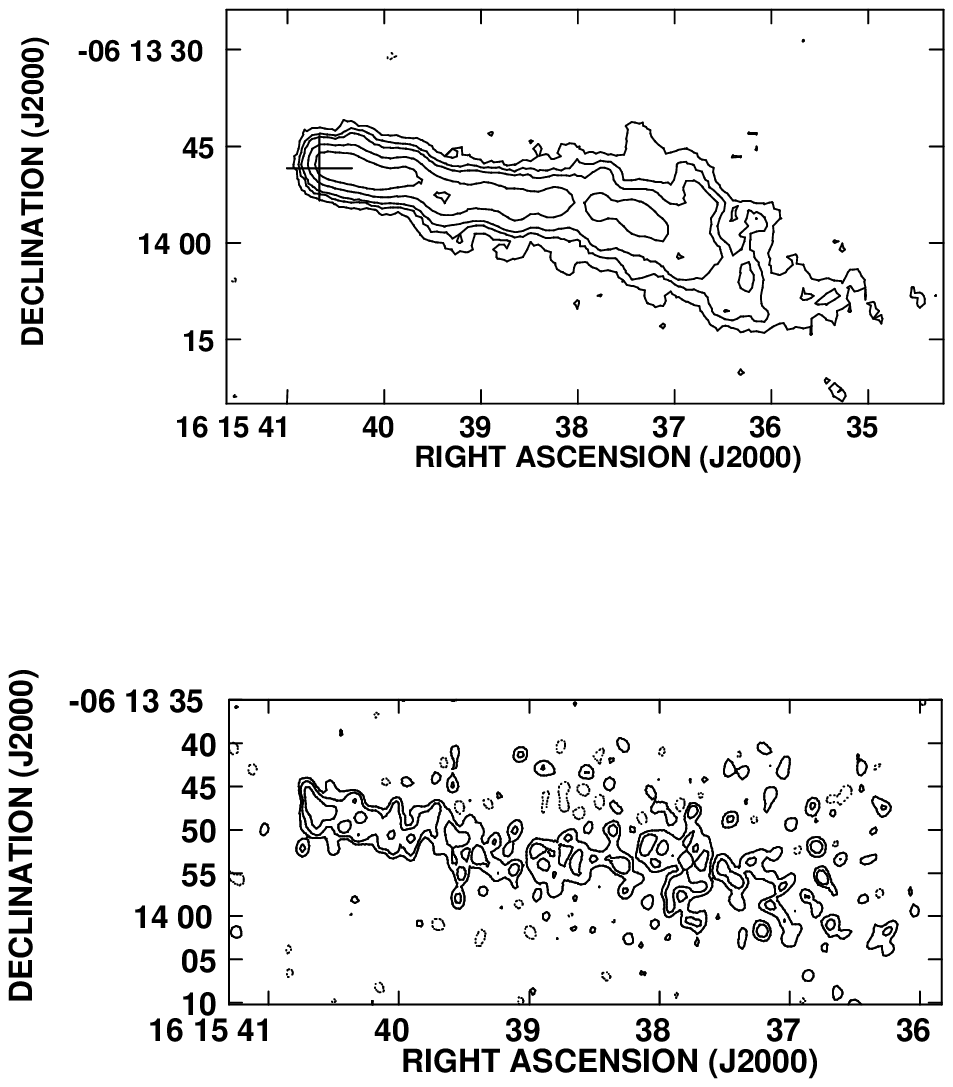}
\vspace{10 cm}
\caption{ High resolution radio images at 20 cm of the tailed radio
 galaxy J1615-062 (T1).
The upper panel shows the map with  resolution of 4\arcsec~ and
$\sigma$ noise level of 0.035 mJy/beam.
Contours are at  --0.1, 0.1, 0.2, 0.3, 0.6, 1  mJy/beam.
The cross indicates the position of a faint galaxy.
The lower panel shows the image with the highest
resolution, of 1.7\arcsec$\times$1.4\arcsec(@PA= --19$^{\circ}$).
The $\sigma$ noise level  is 0.018 mJy/beam.
Contours are at --0.06, 0.06, 0.12, 0.25  mJy/beam.}
\end{figure}

At 6 cm, only the strongest
discrete sources within and around the radio halo are well detected.
The radio halo is not detected because of missing 
short spacings. Also the other diffuse features are completely
resolved at 6 cm.

Since the radio halo is not detected at 6 cm, we 
attempted an estimate of its spectral index by comparing 
the images obtained from the two separate 
VLA frequencies at 20 cm. Although the frequencies are quite close,
the flux densities are slightly different, and the 
estimated spectral index (S$_{\nu} \propto \nu^{-\alpha}$) 
is $\sim$ 1.6$\pm$0.3. This steep spectrum is consistent
with the values obtained in halo sources (e.g. Feretti \& Giovannini 1996).
The map of the spectral index is quite noisy, so it is
impossible to reveal any trend across the radio halo.

\subsection {Tailed radio sources}

\begin{table*}
\caption{Properties of the tailed  radio sources}
\begin{flushleft}
\begin{tabular}{llllll}
\hline
\noalign{\smallskip}
Radiogal. &  Label   &  RA(J2000) &  DEC & S$_{20}$ &  LLS \\
         &           &  h \ \ m \ \ s 
&  $\circ$ \ \ $\prime$ \ \ $\prime \prime$ & mJy  & kpc \\
\noalign{\smallskip}
\hline
\noalign{\smallskip}
J1615-062  & T1 & 16 \ 15 \ 40.6 & --06 \ 13 \ 48 & 34.5$\pm$0.5  & 630 \\
J1615-061  & T2 & 16 \ 15 \ 41.3 & --06 \ 09 \ 08   & 6.0$\pm$0.3 & 170 \\
J1616-062  & T3 & 16 \ 16 \ 05.6 & --06 \ 14 \ 56 & 24.9$\pm$0.4  & 380 \\
\noalign{\smallskip}
\hline
\noalign{\smallskip}
\end{tabular}
\end{flushleft}
Col.1: name, Col. 2: label according to Fig. 2; Col. 3: position of the
radio nucleus; Col. 4: flux density at 20 cm derived from the image
at the lowest resolution; Col 5: largest linear size obtained from the image
at the lowest resolution.
\end{table*}

Three radio sources  in this cluster are found to show a tailed
structure in the high resolution images. 
They are labeled T1, T2 and T3
in Fig. 2,  and are listed in Table 3.
They are all coincident with faint
galaxies on the digitized Palomar Sky Survey images.
Their  
radio images at high resolution are 
presented in Figs. 4, 5 and 6, where a cross indicates
the probable host galaxy. 
The total flux density and size given in Table 3 are 
derived from the maps at lower resolution, since some
structure is missing in the high resolution images. 
The  6 cm images  show the radio nucleus and only the 
beginning of the tail,
confirming the tailed structure of these radiosources.
It is worth noting that all 3 radio galaxies 
have the tails oriented in the same direction (West). 
This could be explained by 
the presence of merger-induced bulk motion of the intergalactic
medium, as suggested by Bliton \etal (1998).

\begin{figure}
\includegraphics{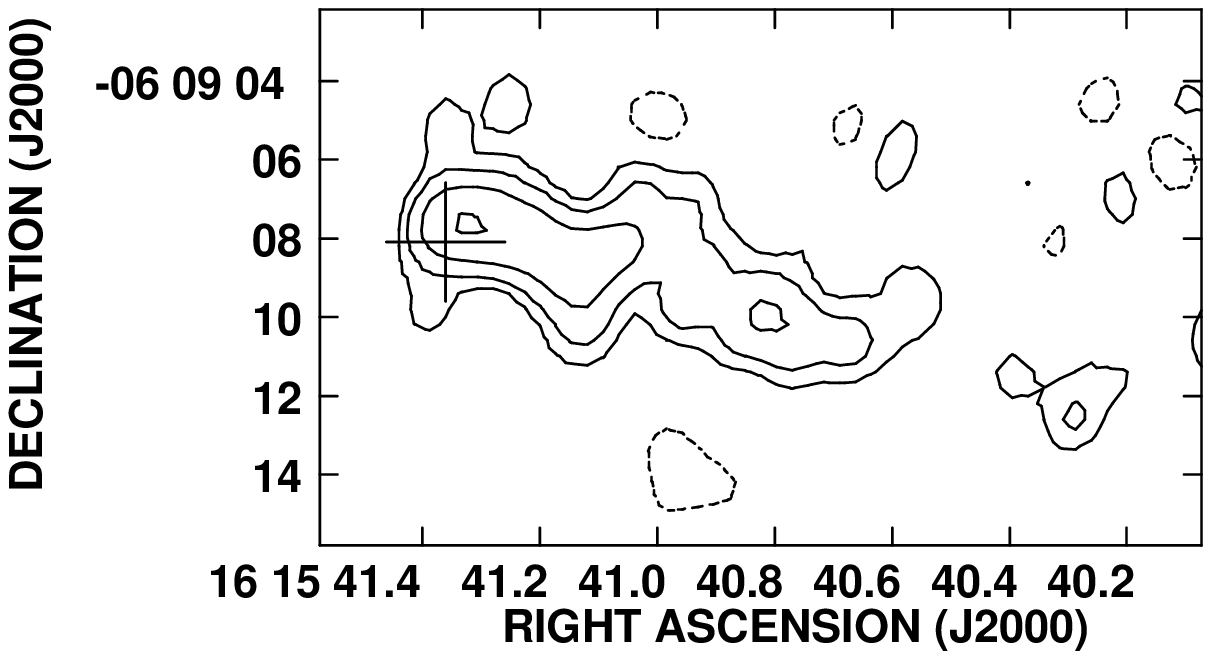}
\vspace{6 cm}
\caption{ Radio map at 20 cm of the radio galaxy J1615-061 (T2)
with resolution of 
1.7\arcsec$\times$1.4\arcsec(@PA= --19$^{\circ}$).
The $\sigma$ noise level in this map is 0.018 mJy/beam.
Contours are at --0.06, 0.06, 0.12, 0.25, 0.5   mJy/beam.
The cross indicates the position of a faint galaxy.
}
\end{figure}

\begin{figure}
\includegraphics{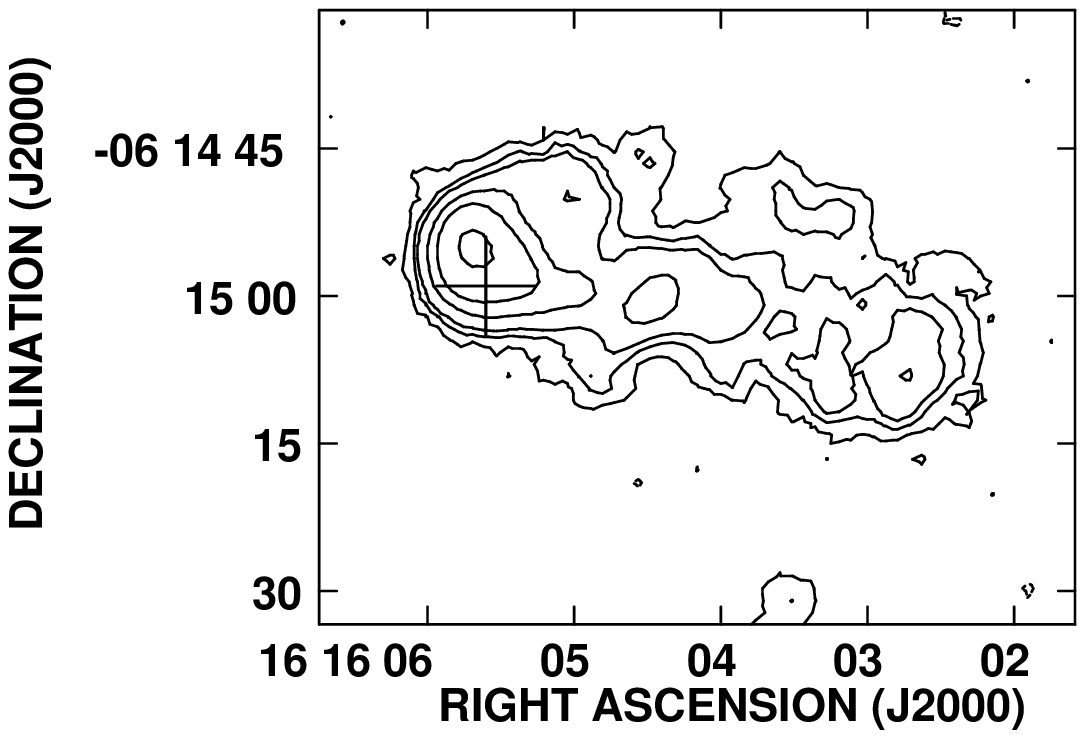}
\vspace{6 cm}
\caption{ Radio map at 20 cm  of the tailed radio galaxy J1616-062 (T3) 
with resolution of 4\arcsec.
The $\sigma$ noise level in this map is 0.035 mJy/beam.
Contours are at --0.1, 0.1, 0.2, 0.3, 0.6, 1, 2  mJy/beam.
The cross indicates the position of a faint galaxy.
}
\end{figure}

\section{Radio versus X-ray comparison}

\begin{figure}
\vspace{6.0cm}    
\caption{Overlay of the radio map (grey scale) onto the X-ray image
(contours). The contour levels are at 0.045, 0.07, 0.11, 0.18,
0.28, 0.45, 0.71, 1.1, 1.8, 2.8 millicts s$^{-1}$ pixel$^{-1}$, with a
pixel size of 15\arcsec $\times$ 15\arcsec.
}
\end{figure}

In Fig. 7 we show the overlay of the radio and X-ray images,
for a morphological comparison. 
The radio image is the same as that presented in Fig. 1. The
X-ray image was obtained with the ROSAT 
Position Sensitive Proportional Counter (PSPC)
with a total exposure time of 12133 sec
(Elbaz  \etal 1995). The image presented here has been reprocessed with a
more accurate determination of the exposure map, and refers 
to the  0.5 - 2 keV energy range, in order to have a lower background.
It has been corrected for the vignetting effect, and smoothed 
with a gaussian of $\sigma$ = 30\arcsec. 
As noted by Elbaz \etal (1995), the cluster X-ray structure is clearly 
non spherical and in particular the shape is
elongated in the E-W direction. There is also some evidence
that the axis of the elliptically shaped distribution
in the central region of the cluster has a different orientation, i.e. 
inclined between the E-W and the NE-SW direction.
These morphological features are strickingly similar to those
found in the radio structure, confirming the connection
between hot and relativistic plasma, found in other clusters
(Deiss \etal 1997, Feretti 1999, Liang \etal 2000).
In addition, we note that there appears to be a compact X-ray source
at the location of the radio source D2.

\begin{figure}
\includegraphics{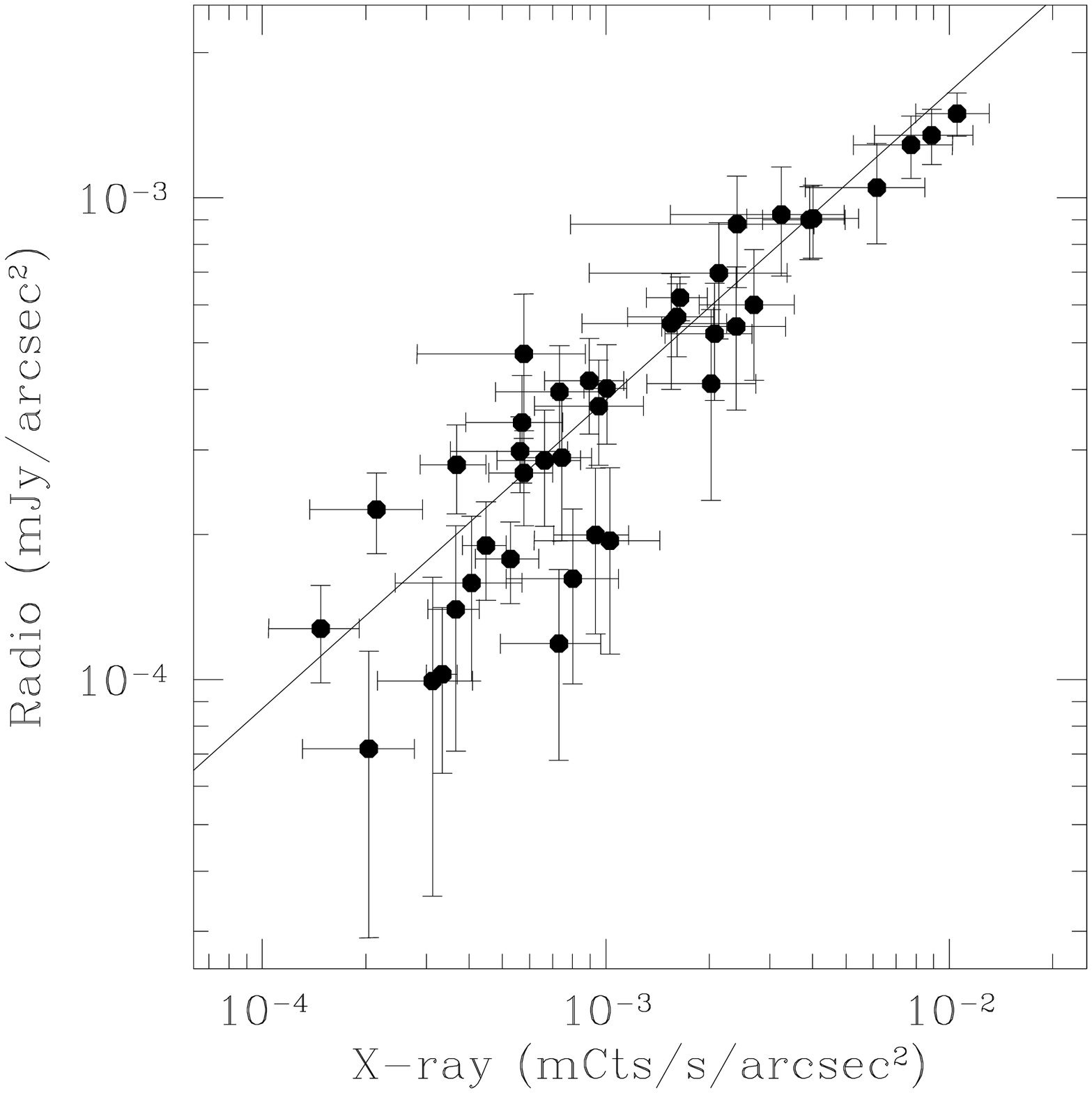}
\vspace{10 cm}
\caption{Plot of the radio brightness versus the X-ray brightness,
obtained in cells of 90\arcsec size. The errors are represented 
by the rms in each cell. The best fit is indicated by a 
continuous line.
}
\end{figure}
 
We have performed a quantitative point-to-point 
comparison of the radio and X-ray brightness, following the same approach 
used by Govoni \etal (2001) 
for the analysis of the clusters Coma, A2255, A2319, A2744.
The radio and X-ray images are suitable for such a study,
having similar resolutions. 
In the X-ray map the constant background was subtracted.
The image obtained after subtraction of discrete
sources and corrected for the primary beam attenuation was used
as a radio map.
The plot of the radio versus the X-ray brightness is
shown in Fig. 8. The close similarity between radio and
X-ray structures is demonstrated by the correlation between these two 
parameters:  a higher X-ray brightness 
is associated with a higher radio brightness. 
 The data was fitted to a power law relation
of the type:
$$ B_{\rm radio} = a B_{\rm Xray}^b$$
With  the radio brightness B$_{\rm radio}$  in mJy arcsec$^{-2}$ and
the X-ray brightness B$_{\rm Xray}$ in millicounts s$^{-1}$ arcsec$^{-2}$,  
the parameters of the best fit are 
b=0.64$\pm$0.05 and  a=0.03$\pm$0.01.
Therefore the correlation indicates that the radial decline of the
non-thermal radio component is slower that that of the thermal one.
Relations with b$<$1 were also  found in the 
clusters Coma and A2319, whereas the correlation between  
the radio and the X-ray
brightness was found to be linear in A2255 and A2744 (b$\sim$1,
Govoni \etal 2001).

\section{BeppoSAX X-ray Observation}

The cluster was observed in February 98 for a total exposure time of 109 ksec. 
Spatially resolved spectroscopy 
derived from the data obtained with the Medium-Energy 
Concentrator Sectrometer (MECS) has been published by Irwin \& Bregman (2000).
The average temperature between 1.65 and 10.5 keV is found to be 
$11\pm 0.6$ keV.
The azimuthally averaged radial temperature profile shows a rather constant 
trend up to a distance of 10\arcmin~ from the cluster center. This result
agrees with the ASCA result of White (2000) but disagrees strongly with the 
result from ASCA and ROSAT by Markevitch \etal (1996), who found a temperature
decrease at large distance.

We have analysed the data from the high energy detector
PDS, operating
in the 15-200 keV energy range, with a field of view of $1.3^{\circ}$
(FWHM, hexagonal) to search for hard X-ray emission, as detected in the
Coma cluster (Fusco-Femiano \etal 1999; Rephaeli, Gruber \& Blanco 1999)
and A2256 (Fusco-Femiano \etal 2000) which both show extended radio regions.
The PDS instrument uses the rocking collimator technique for background
subtraction. The background level is the lowest obtained thus far
with high-energy instruments on board satellites and it is very stable,
thanks to the equatorial orbit. No modeling of the time variation of the
background is required (see Fusco-Femiano \etal 1999). 

The data analysis gives  evidence for hard X-ray emission
at a confidence level of $\sim 3.4\sigma$, with a  
count rate of 0.219$\pm$0.065 c s$^{-1}$ in the PDS 
energy range. The data is well fitted by a thermal component at the
average cluster temperature of 11 keV. For the non-thermal emission,
an  X-ray flux upper
limit  of $\sim 5.6\times 10^{-12}$ erg cm$^{-2}$ s$^{-1}$ in the 20-80 keV
energy range is derived.
This upper limit is about 4 and 2 times lower
than the non-thermal flux measurements obtained
in the same energy range for 
Coma (Fusco-Femiano \etal 1999) and
A2256 (Fusco-Femiano \etal 2000), respectively.

\section{Discussion}

\subsection{Halo origin}

The diffuse halo in  A2163 is more powerful and more extended than the
prototype halo source in the Coma cluster, Coma C. Like Coma C,
it is rather regular in shape.
The properties of radio halos and relics in clusters have been
recently reviewed by Feretti (2000) and Giovannini \etal (2000).
The formation of radio halos has been found to be strictly related  
to the  X-ray properties of the host clusters and to the presence
of cluster merger processes. In particular, the
percentage of clusters showing radio halos  is much higher in 
a sample of X-ray
luminous clusters ($\sim$30\% for L$_{\rm X-ROSAT}$ $>$ 10$^{45}$ erg s$^{-1}$)
than in a complete cluster sample ($\sim$5\%). 
The clusters hosting a diffuse source have a
significantly higher X-ray luminosity than
clusters without a diffuse source.
Moreover, the radio properties of halos are linked to the properties
of the X-ray emitting gas through correlations (Feretti 2000, Liang \etal
2000): the halo monochromatic
radio power correlates with the X-ray luminosity, the X-ray temperature and
the cluster mass. A correlation is also found between the largest
radio size of halos and relics and the cluster X-ray luminosity,
with  more X-ray luminous clusters hosting larger  sources.

The properties of the radio halo in A2163 are fully consistent with the 
above suggestions. Indeed, this cluster is very hot and massive, and 
hosts the most powerful and extended  halo known. 
From the temperature maps obtained with ASCA 
and recently with Chandra, Markevitch \etal (1996) and 
Markevitch \etal (2000)  inferred that the 
cluster inner region is 
in a highly nonrelaxed state, with shocks or streams 
of shock-heated gas, probably due to a recent merger. This is also confirmed
by Squires \etal (1997) from   
the study of the distribution of  cluster light, thermal gas 
and dark matter.  
The merger process in this cluster is likely to provide
the energy necessary for particle reacceleration, according to a model
suggested by Brunetti \etal (2001) for the Coma cluster. 

The radio structure of this halo  shows close similarity to 
the structure of the X-ray emitting gas. This is evident from the
overlay between radio and X-ray images, and from the 
correlation between radio and X-ray brightness.
The connection between the hot and relativistic plasma could
still reflect the connection between the maintenance of the radio halo
and the energy provided by the cluster merger,  and, when available for
a large sample of clusters, could  be
used to discriminate between models of halo formation 
(see Govoni \etal 2001). 

The presence of a relic in this cluster is very interesting. It is
located at about 2.2 Mpc from the cluster center and is oriented
roughly perpendicular to the cluster radial direction, like the other
well know relics. The particles
radiating in the relic could be reaccelerated in a shock produced
by the cluster merger, as in A3667 (Roettiger \etal 1999).
Unlike the cluster A3667, which has no radio halo at  the
center,
the existence in the same  cluster of both a central radio halo
and a peripheral relic is not uncommon (Feretti 1999).
This points in  favor of a common origin for halos and
relics, which could  be investigated when detailed
data on the cluster merger state and evolution will be available
from the new generation X-ray telescopes.

\subsection{Inverse Compton emission}

Non-thermal hard X-ray radiation is expected in galaxy clusters
with radio halos, because of inverse Compton (IC) scattering
by the radio emitting relativistic electrons with the 
CMB photons. The alternative interpretations
to the IC model  proposed in the literature  
for the non-thermal emission detected in Coma 
(En{\ss}lin, Lieu \& Biermann 1999; Sarazin \& Kempner 2000; Blasi \&
Colafrancesco 1999; Dogiel 2000; Blasi 2000)
show serious difficulties, as recently pointed out by Petrosian (2001). 

From the PDS data analysis we derive  evidence for hard X-ray emission
at a confidence level of
$\sim 3.4\sigma$. This can be 
accounted for by the cluster thermal emission, whereas for the non-thermal
emission only an upper limit is derived.
The detection of non-thermal emission in A2163 appears problematic for the
great distance and high gas temperature of this cluster.

Within the framework of the IC model, it is
possible to derive, only using observables (see Rephaeli 1979), 
 a lower limit to the volume-averaged intracluster
magnetic field combining the synchrotron radio flux with the X-ray
flux upper limit for the non-thermal component (see Sect. 5).
Using the radio data reported in Table 2 for the halo, we
obtain B \gtsim 0.28 $\mu G$. Assuming a radius of
$\sim$1.4 Mpc for the radio emitting region, at the distance of the cluster,
we derive an upper limit to the electron energy density, $\rho_e$, of
$\sim 6.5\times 10^{-14}$ erg cm$^{-3}$.
The figures obtained in this way are
consistent with the equipartition
values (see Table 2; note that the minimum energy density in the table 
includes the contributions of both the radiating particles and the magnetic
field). The derived limits  are not sufficient to establish
whether the radio halo is in equipartition conditions
or not.

\begin{acknowledgements}

We are indebted to
 Monique Arnaud for providing the X-ray ROSAT image, and
for helpful discussions. The point-to-point radio-X-ray 
comparison was performed with the Synage++ package developed
by Matteo Murgia. We thank Barry O'Connell for comments on
the manuscript.
The National Radio Astronomy Observatory
is operated by Associated Universities, Inc., under contract with the 
National Science Foundation.

\end{acknowledgements}

\end{document}